\documentclass[letterpaper]{article} 
\usepackage{aaai25}  
\usepackage{times}  
\usepackage{helvet}  
\usepackage{courier}  
\usepackage[hyphens]{url}  
\usepackage{graphicx} 
\usepackage{natbib}  
\usepackage{caption} 
\usepackage{algorithm}
\usepackage{algorithmic}

\usepackage{amsmath}
\usepackage{amssymb}
\usepackage{booktabs}
\usepackage{graphicx}
\usepackage{threeparttable}
\usepackage{xcolor}
\usepackage{soul}

\usepackage{newfloat}
\usepackage{listings}
\DeclareCaptionStyle{ruled}{labelfont=normalfont,labelsep=colon,strut=off} 
\floatstyle{ruled}
\newfloat{listing}{tb}{lst}{}
\floatname{listing}{Listing}
\title{Enabling Rapid Shared Human-AI Mental Model Alignment via the After-Action Review}
\author {
    Edward Gu*,
    Ho Chit Siu*,
    Melanie Platt, 
    Isabelle Hurley, 
    Jaime Peña, 
    Rohan Paleja
}
\affiliations{
    Lincoln Laboratory\\
    Massachusetts Institute of Technology \\    
    Lexington, MA 02142 USA\\
    \{edward.gu, hochit.siu, melanie.platt, isabelle.hurley, jaime.pena, rohan.paleja\}@ll.mit.edu
}

\begin{document}

\maketitle

%

\begin{abstract}

In this work, we present two novel contributions toward improving research in human-machine teaming (HMT): 1) a Minecraft testbed to accelerate testing and deployment of collaborative AI agents and 2) a tool to allow users to revisit and analyze behaviors within an HMT episode to facilitate shared mental model development. 
Our browser-based Minecraft testbed allows for rapid testing of collaborative agents in a continuous-space, real-time, partially-observable environment with real humans without cumbersome setup typical to human-AI interaction user studies.
As Minecraft has an extensive player base and a rich ecosystem of pre-built AI agents, we hope this contribution can help to facilitate research quickly in the design of new collaborative agents and in understanding different human factors within HMT. Our mental model alignment tool facilitates user-led post-mission analysis by including video displays of first-person perspectives of the team members (i.e., the human and AI) that can be replayed, and a chat interface that leverages GPT-4 to provide answers to various queries regarding the AI's experiences and model details. 

\end{abstract}
\section{Introduction}\label{sec:introduction}



Coordination across multiple agents has been a long-studied problem in Artificial Intelligence (AI) across a variety of games, including Google Research Football, Hide-and-Seek, and Starcraft \cite{kurach2020google,Baker2019EmergentTU,vinyals2017starcraft}. Various facets of this problem have been explored, including partial observability, limited (or lack of) communication between agents, and heterogeneous capabilities across teammates. Solutions in this area have surrounded approaches that build an implicit \textit{shared mental model} across teammates \cite{Rashid2018QMIXMV}, thereby aligning teammate behavior to achieve a common objective. Formally, a mental model is an agent's internal representation of itself in relation to a task, the environment of the task, and any other agents in the environment \cite{staggers1993mental}. Forming an accurate mental model of one's teammates is an important part of working together in a collaborative task, and a team's collective understanding of a task and how to work together in it is considered a ``shared'' mental model \cite{Mathieu2000TheIO}.

More recently, AI researchers have begun to study the development of collaborative agents that can team with humans, seeking to augment and empower humans. However, several works have shown that introducing AI agents is not as simple as deploying a high-performing trained agent \cite{siu2021evaluation}. As humans, when teaming with other humans, require training to build a shared mental model to collaborate well with teammates, humans need a similar process when teaming with agents. 
Explainable AI (xAI) techniques are one pathway to help humans develop such a mental model as they elucidate otherwise hidden decision processes internal to agents \cite{Sanneman2023UnderstandingOR}. However, such techniques are not often augmented with a pathway (e.g., interface) to allow for a human to understand and stratify explanation content to allow for building a global understanding of a decision-making agent.

We propose a different formalism to help humans and their collaborative AI agents build a model model, the \textit{After-Action Explanation} (AAE), based on the After-Action Review (AAR) process. The \textit{after action review}, also known as the instructional debrief, is a process by which human teams can develop a shared mental model of the events, decisions, and thought processes that occurred during a recent team task to improve future performance and teaming. This process is most commonly used in a military context, where members of a team execute a mission, then review the mission on their own, and then, finally, review the mission as a group \cite{deptula2012fundamentals}. In particular, focus is given to identifying and correcting errors that occurred during the course of the task, 
typically categorizing the cause of an error in terms of perception, decision-making, or execution. The debrief process typically involves reviewing video and communication recordings of the mission, along with the history of discrete events, such as changes in objectives, or new information gained.

The after-action review is also used prominently in medical education, particularly after simulation activities \cite{abulebda2019debriefing,cheng2014debriefing}. Medical instructional debriefs follow a somewhat different and more varied set of frameworks from the military context, though the overall goals of reflection, correction, and subsequent improvement remain.

In this work, we will refer to the human-only process as after-action review, and the human-AI explanatory process as after-action explanation. We leverage a large language model (LLM) to facilitate the AAE, allowing for humans to input free-form queries and follow-up questions and receive adaptive responses tuned via prompting and user input into the LLM. The LLM draws upon agent experience (observations and actions) and agent model details that have been annotated into text files to produce AAEs.

An essential component of the AAR/AAE process is the ability to conduct multiple iterations of teaming followed by review. For HMT, this would involve conducting extended or multi-session experiments with humans. As this type of deployment is resource-intensive, we require flexible, lightweight HMT testbeds that can simulate complex collaborative tasks. Currently, the most widely known testbed for human-machine collaboration is the Overcooked-AI testbed \cite{carroll2019utility}. While easy-to-use and applicable to a broad range of problems, this domain is relatively low-dimensional. To allow for easy deployment of collaborative agents to a higher-fidelity testbed that supports an abundance of tasks, we create a \textcolor{black}{browser-based} HMT testbed in Minecraft, allowing for rapid testing of and human-subjects research with agent models with humans without cumbersome setup.

\begin{figure*}
    \centering
    \includegraphics[width=0.8\linewidth]{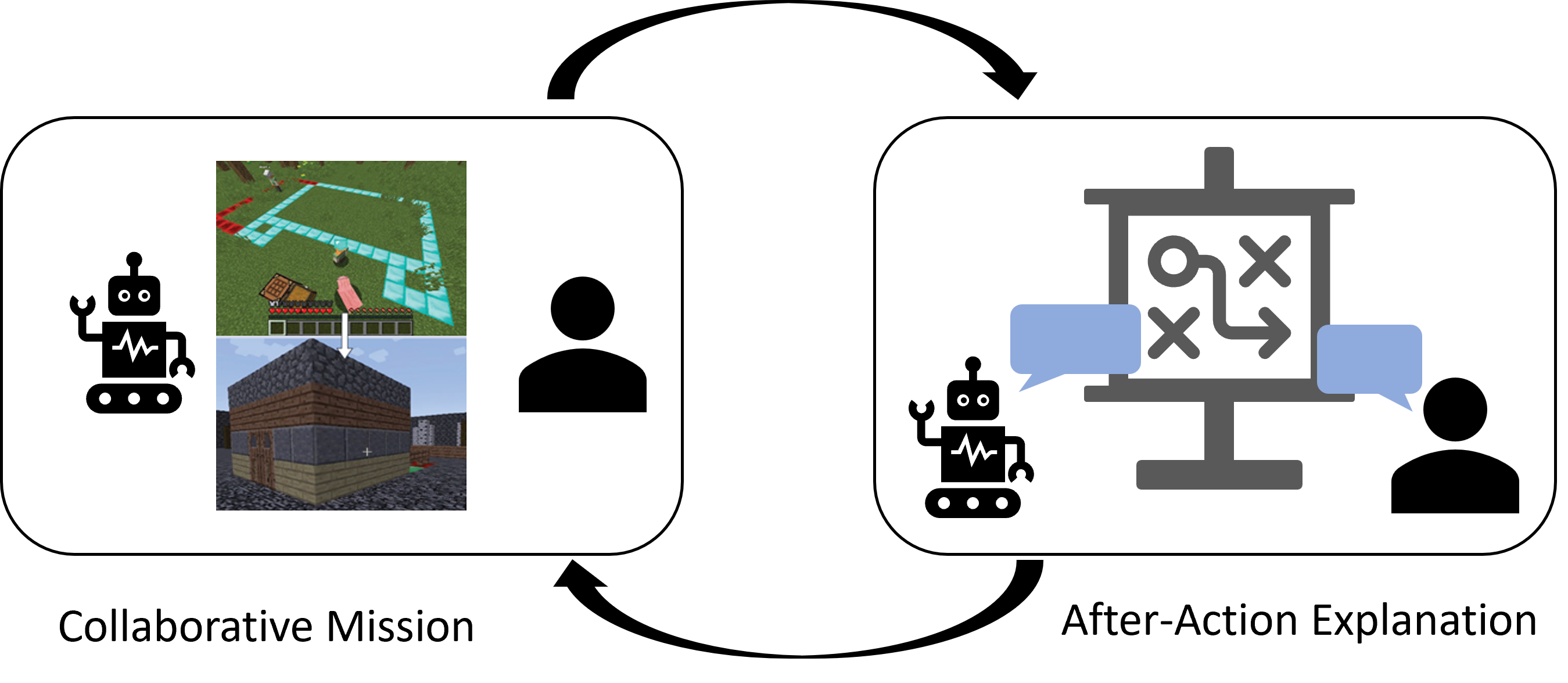}
    \caption{After-Action Explanation (AAE) workflow for human-AI teaming in Minecraft, modeled after the After-Action Review procedure. The cycle begins with humans and AI agents performing a collaborative building task together. Following task completion, engage in a group debrief session where they can discuss and understand the mission outcomes, agent behaviors, and decision-making processes through the AAE tool, facilitating the development of shared mental models between human and AI teammates.}
    \label{fig:aae_workflow}
\end{figure*}

\noindent We summarize our contributions as follows:
\begin{itemize}
    \item A human-machine teaming (HMT) testbed in Minecraft to accelerate research in human-machine collaboration. 
    \item An After-Action Explanation (AAE) Tool (Figure \ref{fig:aae_workflow}) to facilitate after-action review post-brief mission evaluation. 
\end{itemize}

\section{Related Work}
In this section, we present related work in Shared Mental Models, xAI for Collaborative Agents, LLMs as an Interface for Explainability, and Human-Machine Teaming Testbeds.

\paragraph{Shared Mental Models}
Human-human teaming behavioral research has demonstrated that teams are more effective in coordinating their behavior when they can account for each other’s intentions, beliefs, and behavior (i.e., have a shared mental model) \cite{Mathieu2000TheIO,DeChurch2010TheCU}. While active communication can help humans convey information and achieve shared goals \cite{Salas1992TowardAU}, communication is not always possible and can often degrade team performance if not used judiciously. In this work, we are focused on an offline technique for building shared mental models, the After-Action Review. AARs are used across a variety of fields, ranging from military applications \cite{deptula2012fundamentals} to medicine \cite{abulebda2019debriefing} and business operations \cite{Aguinis2020MethodologicalPI}. AAR techniques aim to help operators understand deviations from expected mission outcomes and align behavior toward continual team improvement.

\paragraph{Explainable AI for Collaborative Agents}
A promising direction utilizes xAI for developing shared mental models in human-robot teams. xAI focuses on making AI systems understandable and interpretable to humans \cite{linardatos_explainable_2021,dwivedi_explainable_2023}. Diverse explanation types have been shown to elucidate robot behavior for human teammates, such as presenting adverbial cues for correcting false beliefs \cite{briggs2011facilitating} and decision-tree policy descriptions for increasing situational awareness \cite{paleja2021utility}. Visual explanations, including saliency maps \cite{anderson2020mental} and planning diagrams \cite{wortham2017improving}, as well as propositional explanations \cite{Sreedharan2018hierarchical}, demonstrate improved accuracy in human mental models of robots on various measures. While these works indicate that explanations generally improve human mental models of the robot, they come with tradeoffs, such as increased cognitive load for the human \cite{anderson2020mental} and have different effects based on human expertise \cite{Sreedharan2018hierarchical}. Other works have begun to employ natural language, seeking to allow for more straightforward human interpretation. \citet{rosenthal2016verbalization} employ ``verbalizations'' of a robot's navigation plans to convey its intentions to the human teammate with different levels of abstraction based on user preferences. \citet{hayes2017improving} develop a question-answering system enabling robots to respond to human teammates’ queries with natural language explanations describing their behavior. \citet{schraggen2020trusting} use both causal and intentional explanations to communicate the decision-making of an autonomous vehicle, showing in a user study that explanations overall assisted in increased participant understanding of the system and establishing appropriate trust. These studies are an encouraging approach for developing accurate shared mental models, but explanations are limited to pre-defined or finite phrase combinations. To enhance the understanding of robot behavior and beliefs among human teammates working in close collaboration, more expansive and flexible communication is required between the human and robot.





\paragraph{LLMs as an Interface for Explainability}
Large Language Models (LLMs) such as GPT-4 \cite{openai2024gpt4technicalreport} have advanced capability for generating human-like conversational language and are a promising direction for translating explanations to be more understandable to human end-users. Recent work has shown LLMs can generate explanations of their own reasoning when problem solving \cite{wei2023chain,huang2023large}, as well as explanations of other model’s decisions (e.g., image classification \cite{mozannar2023effective, tursun2023selfexplainability}). 
Although limited, new research is emerging that utilizes LLMs to explain robot behavior. \citet{zhang2023explaining} utilized LLMs to explain robot behavior by passing behavioral representations of the state and action space as context. Their work found that users preferred LLM-generated explanations over rule-based explanations and found the chat feature helpful when the robot's policy was suboptimal. \citet{gonzalez2023usinglarge} also leverage an LLM to explain an autonomous robot, inputting various logs of behavior as context, and found significant challenges with the accuracy of the produced explanations. In a similar task setting, \citet{SobrnHidalgo2024ExplainingAE} leverage Retrieval-Augmented Generation (RAG) systems to produce explanations regarding log files detailing the robot's behavior. In a user study, the authors found that while the model generated clear, understandable explanations, they were not consistently accurate. 
\textcolor{black}{Our contribution is unique from these works in that they consider robots acting alone to complete a task, while we are interested in humans and robots acting collaboratively and the influence of explanations on their shared mental model. We further leverage a real-world explanatory process used in team operations, rather than generically providing explanations.}

\paragraph{Human-Machine Teaming Testbeds}

There are several existing human-AI collaboration platforms. Gym-Cooking is a gridworld cooking game inspired by the game Overcooked \cite{wu2021too}. The Hanabi card game challenge \cite{bard2020hanabi} also spurred the creation of a human-interactive environment \cite{lerer2020improving}, creating an interesting multi-agent, turn-based, game with partial observability. 
The HMT testbed that we create is based on Minecraft. In contrast to the above platforms, Minecraft is a continuous-space, real-time, partially-observable game with highly-customizable objectives. The partial observability in Minecraft is different from most other human-AI environments as it stems primarily from the first-person point of view of the agents rather than explicit information hiding (Hanabi) or occlusions in an otherwise global game view, thus mimicking the kind of partial observability encountered by teams of humans and mobile robots. Additionally, apart from the platform itself, Minecraft is a game for which many AI agents have already been built \cite{Gray2019CraftAssistAF}, and for which there is a large and active human player base. The latter two community-based factors are extremely important --- when using Minecraft, researchers can draw from an existing and diverse set of pre-built agents, rather than training from scratch or relying on sample agents from one or two developers. Furthermore, unlike other collaborative games with small user bases (larger population of unfamiliar users) where human-AI experiments likely exhibit a significant learning effect early on, the associated large user base in Minecraft allows for ease-of-recruiting for human participants who are already experts in the mechanics of Minecraft. While Minecraft does have a prior platform supporting research \cite{Johnson2016TheMP}, which was extended in \citet{paleja2021utility}, this version of the testbed is cumbersome to set up and not scalable. \textcolor{black}{In our proposed HMT testbed (Section \ref{sec:system}), we take steps toward building a testbed where human-subject study participants can log into a server on their own computer and play with a collaborative agent with little to no setup.}



\section{Flexible Human-Machine Teaming Testbed}
\label{sec:system}
In this section, we provide details regarding our HMT testbed, carefully detailing the different software blocks that were integrated and the features our testbed supports. 

\subsection{Architecture}

\begin{figure*}
    \centering
    \includegraphics[width=\linewidth]{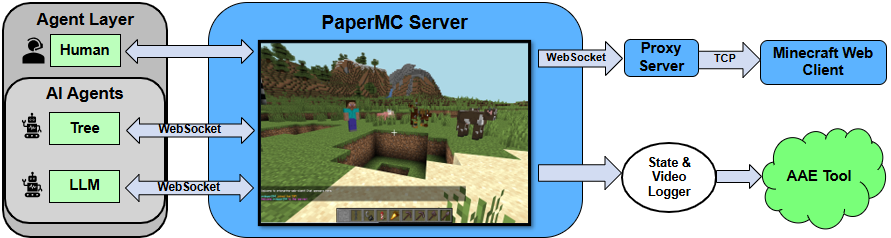}
    \caption{HMT Minecraft Infrastructure: The system architecture consists of four main components (1) an Agent Layer supporting both human players and AI agents (e.g. decision tree agent, LLM agent), (2) a PaperMC server that manages the game environment and agent interactions, (3) a web-based Minecraft client, and (4) an After-Action Explanation tool. A proxy server is used to connect the PaperMC server with the web client. All components communicate through WebSocket connections, except for the TCP connection between the proxy server and web client.}
    \label{fig:hmt_system}
\end{figure*}

Our system infrastructure is composed of several components:

\begin{itemize}
    \item Minecraft Server: A custom offline PaperMC server (https://papermc.io/) that manages the game world, handles player authentication, and coordinates multiplayer interactions. It maintains world state and processes game events.
    \item Minecraft Web Client: An offline browser-based interface (mcraft.fun) powered by PrismarineJS that allows users to connect to and interact with the Minecraft server without installing a desktop client.
    \item Proxy Server: A server that mediates communication between the web client and Minecraft server, handling protocol translation and ensuring secure data transmission.
    \item WebSocket: A communication protocol that enables real-time, bidirectional data exchange between the Minecraft server, AI agents, and the AAE tool, transmitting player positions, inventory states, and actions.
\end{itemize}


The integration of these components can be seen in Figure \ref{fig:hmt_system}. Importantly, combining these capabilities allows us to connect varying AI agents that can function over various levels of abstraction, design a variety of tasks that can vary in complexity and average completion time, customize and annotate data received through HMT gameplay, allow for HMT experimentation with minimal setup (and commute time), and allow for the integration of different web wrappers that can allow for conducting surveys or receiving After-Action Explanations.

\paragraph{World Environment Setup}
Out of many potential Minecraft servers, we use the highly popular PaperMC server with a custom Bukkit plugin that initializes a controlled Minecraft environment for our human-AI collaboration tasks. The server environment is set to be an empty flat world that is then populated with predefined locations of blocks. In our current task setting, we focus on collaborative construction of a house, similar to \citet{paleja2021utility}. We mark the ground with different block types that relate to a floor plan,
introduce resource towers containing various types of blocks needed to build a house,
and utility blocks such as a crafting table, which allows for constructing tools for faster resource collection, and a storage chest, which facilitates resource sharing between the human and the AI. \textcolor{black}{While we focus on collaborative building, we provide instructions for developers to design their own tasks.}

\paragraph{Player-AI Management and Communication}
Alongside world generation, the Bukkit plugin handles connections to AI agents and connections to other web programs, including the AAE tool and survey platforms. We have control over the spawn points of the human player as well as any AI agent, as well as environmental parameters that control scene lighting or adversary generation.
A WebSocket server system is set up to handle communication to and from the server. The WebSocket streams real-time human player data, including position coordinates, inventory contents, item held, block it is looking at, and behavior state to the AI agent(s), and receives similar state updates and action requests from the AI agent(s). The plugin is also able to execute AI agent commands, such as crafting a pickaxe or storing items in a chest. \textcolor{black}{A detailed step-by-step process is provided in our codebase for developers to test their own custom human-aware AI agents.}


\paragraph{Video and State Logging}

Our Minecraft HMT testbed captures key information about the environment and all agents. In our work, this information is critical for the AAE tool which must explain robot behavior to assist in mental model alignment. All human, AI, and world states are logged in a JSON format when the game begins (when both the human and the AI agent join the world) and logging ends when the mission ends in success or failure. In our task, success is determined when the house has been completely built before 15 minutes have elapsed, according to the specifications provided, and failure is indicated by incomplete completion of the house by the allotted time. Changes to the agent's position, inventory, action (e.g. mining, crafting, idle), and other state parameters are synchronously logged by the system. 
\textcolor{black}{We focus on logging events as opposed to observation-action pairs at a pre-specified frequency to maintain an easier-to-analyze data file that the AAE system can ingest.} First-person video from both the human and the AI's perspectives, as well as a top-down view of the world, are automatically recorded during the mission. These visualizations are important for a human to successfully perform an AAE and revisit different periods of a teaming interaction. 

\subsection{Minecraft AI Agents}

Within our HMT testbed, we provide two AI implementations that can assist with collaborative house building: 1) a white-box decision tree AI and 2) an LLM-enabled AI agent. Our AI agents are implemented within Mineflayer, a widely used JavaScript API for Minecraft AI agent development. Mineflayer enables key AI agent capabilities such as navigation (A* search for efficient path-finding to a target location while avoiding obstacles), resource mining, tool crafting, and storing resources in the chest. The decision tree AI's policy is based on \citet{paleja2021utility} and is written using the Mineflayer API. The LLM-enabled AI agent is directly pulled from \citet{mindcraft2023} and calls basic skills already written in the Mineflayer API, utilizing an LLM to translate human commands into sequenced skill calls. For either of these AI agents, multiple AI can be spawned and carry out their policies simultaneously alongside the human agent. However, in our work, we focus on a single AI agent collaborating with a single human. We provide brief detail below regarding the behavior of each AI below.


\paragraph{Decision Tree AI}
We adapt our decision tree AI agent from \citet{paleja2021utility}, where the AI agent operates on a five-phase decision tree system that is conditioned on the human's state. Switching between the different phases of the decision tree is dependent upon the completion percentage of the house-building task. The initial phase is always one, and the AI agent shifts to higher-numbered phases as the house completion score increases to predefined thresholds. A human behavior inference system tracks the human's proximity to key locations, identifies current activities based on position, and adapts the AI's behavior to complement human actions. It is important to note that this AI agent is constrained so that it is unable to place blocks down, a capability reserved only for the human agent, further reinforcing that the AI agent needs to actively collaborate with the human in order to build the house.
Additional tree phases or nodes may be readily added to handle other independent variables, augmenting the AI's capabilities to handle a wider variety of tasks. Examples include defending against adversaries (e.g. creepers, slime), interacting with neutral non-player characters (e.g. villagers, animals), and \textcolor{black}{communicating via the chat interface}. Phase information, active decision tree branches, and the selected decision node are saved alongside state information at each event.

\paragraph{LLM-enabled AI}
For the LLM-enabled AI, we pull MINDcraft’s implementation of the “Andy” AI agent that is designed to follow natural language instruction from a human player \cite{mindcraft2023}. This AI agent uses ChatGPT \cite{chatgpt} to call commands from a library of skills and actions written in Mineflayer. When asked to complete an instruction that the AI agent does not know how to do, it will dynamically write new code in Mineflayer to do so. Finally, after the instruction and resultant action have been decided, the AI responds back to the human via the Minecraft chat with a conversational response and the command that it is going to execute. The conversation history between the human and AI agent is saved at the end of each session, which we have amended to include a timestamp. This AI has the same capabilities as the human agent, including the ability to place blocks down. Thus, there are two modalities of collaboration that can take place. In the first, the human acts as a commander, and the agent acts as a worker that interprets the human's commands and executes the task. In the second, the AI agent is given a predefined context file (or at the beginning of the mission) that includes high-level goals and constraints. In this case, the AI agent reacts to the human and acts as an assistant for the human, much like the decision tree AI agent.

 \noindent\textbf{Dislclaimer regarding the LLM-enabled AI} The issue with current LLM AI agents is that they are imperfect and cannot handle high-level commands like ``build a house using x,y,z resources'' or write policy code to \textcolor{black}{infer} human behavior or align with implied human goals during collaboration tasks. While MINDcraft has the capability to write code, the user will likely have to go through the code generated by the LLM AI agent and painstakingly find out where the errors in the code are and manually fix them. 
 

\textit{ For both agents, the AAE tool can be used to explain AI behavior, and possibly resolve any discrepancies between expected AI behavior and the actual AI behavior. }

\section{After-Action Explanation Tool}

\begin{figure*}
    \centering
    \includegraphics[width=0.8\linewidth]{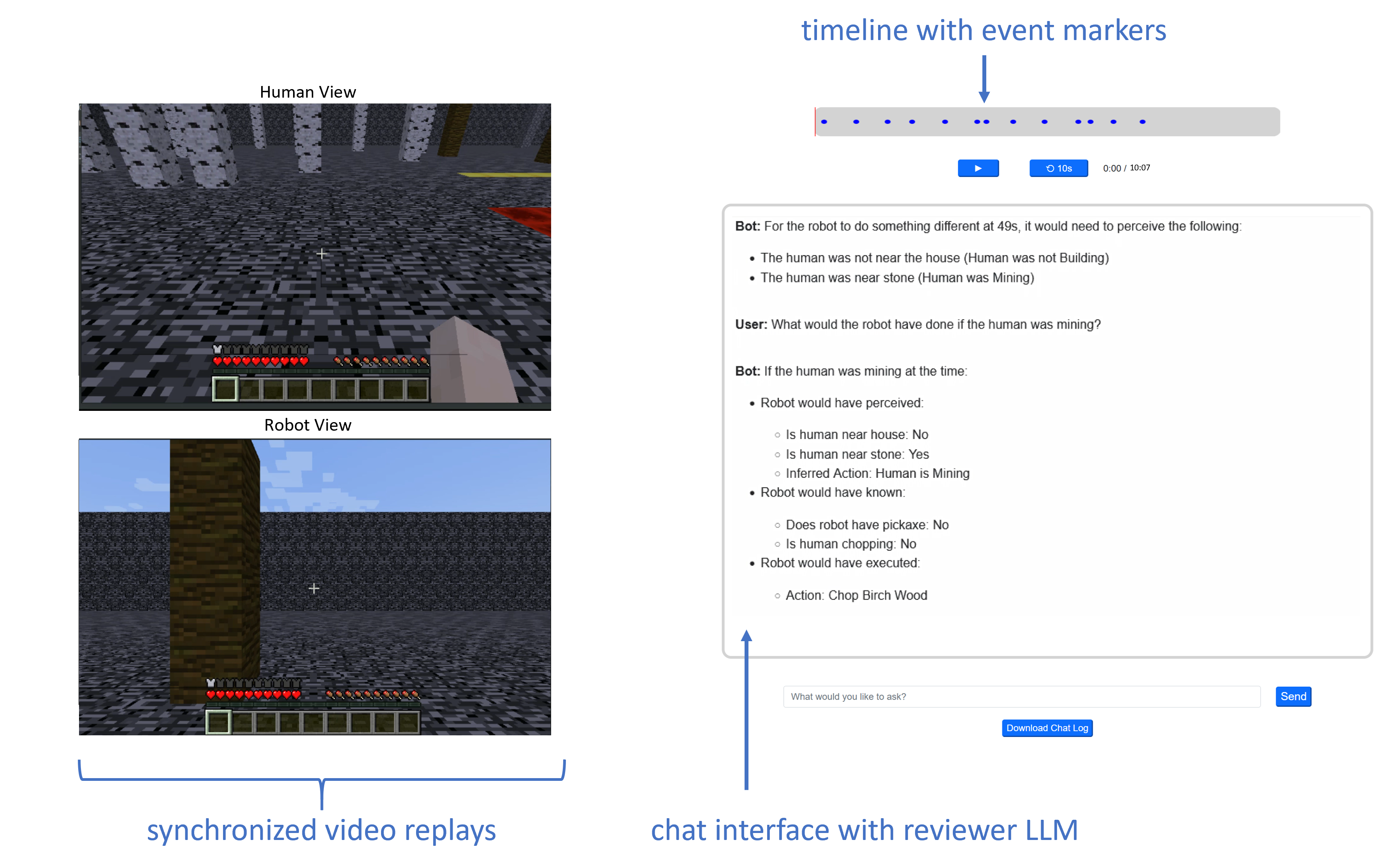}
    \caption{After-action explanation interface. Human and AI agent points of view are show in synchronized video on the left. Timeline with events marked in blue is shown on top right, and the LLM chat interface is on the right. The mission context and timeline files are loaded automatically.}
    \label{fig:aae_tool}
\end{figure*}

The After-Action Explanation Tool is made to emulate an AAR. This tool is a web application that consists of four integrated components 1) synchronized video replay of human and AI behavior, 2) a mission context document, 3) a mission timeline file, and 4) a chat interface. The full tool can be visualized in Figure \ref{fig:aae_tool}.

\paragraph{Synchronized Video Replay} The replay element shows two videos from the perspective of the human and AI agent, respectively, during the course of the mission. These videos are synchronized on a single visual timeline, allowing the human participant to replay the episode and identify key points that they may have questions about. We augment the play bar with markers for specified events (in our case, decision points) so that the user can more easily navigate to periods where the AI behavior shifted. Such a feature is critical for longer HMT missions as the larger volume of interaction data can overwhelm the human operator, making it difficult to recall key moments and extract critical insights independently. 

\paragraph{Mission Context} The mission context document is a text file that provides any relevant information known about the mission and the agents before the mission is run. In typical usage, it describes items such as the purpose and phases of the mission, the number of agents involved, which agents are humans vs AI, and what the capabilities and decision-making processes are of each agent. Information may also be provided in this document about how the chat interface should respond to queries. The context document can readily be augmented with documentation generated with the AI agent (e.g., technical papers or READMEs), allow for human participants to receive in-depth answer regarding the behavior of an AI.

\paragraph{Mission Timeline} The mission timeline file is a JSON file that describes the course of events that occurred in a particular mission, including any public, private, or world-state knowledge present. The root element is an array of objects, each of which represents an ``event.'' Events are defined by two keys, a \textit{timestamp} for the event, and an \textit{action}, which itself is an object containing information about the agents. The events in the JSON are used to mark locations on the visual timeline that is used to control video replay. In typical usage, the timeline can contain information such as agent states, actions, and beliefs, as well as world state information such as task progression. This document is critical for the LLM to accurately answer questions regarding the AI's decision-making. In Section \ref{sec:system}, we describe how we carefully decided how state-observation information should be inputted into the AAE tool for different agents.

\paragraph{Chat Interface} The chat interface is connected to a large language model and is used alongside the video replays to mediate the user's examination of the mission. The mission context and timeline files are provided as part of the prompt whenever a query is entered, allowing the LLM access to both the history of the conversation and relevant mission information. The LLM does not take visual input from the video replays but rather relies on the information from the provided files. However, to facilitate interaction, the video timestamp currently being displayed is included as part of the prompt whenever a query is made so that the LLM has context for questions about what the user is seeing ``at the moment.''

\paragraph{Example Usage}
Given the state log of the game play, the AAE tool aids in the human's search for any discrepancies between what the agent is supposed to be doing and what it is actually doing. 
Suppose during review, the human observes in the videos displayed on the AAE interface that the decision tree agent is continuing to chop more wood logs even though the human already has sufficient wood logs in its inventory to build the entire house. The human queries the LLM regarding what phase of the decision tree the AI agent was in during that time to gain a better understanding of why the AI agent continued performing an action that was unnecessary. From this information, the human can better understand how to work with this decision tree policy in the next iteration of game-play. 

\section{Example User Study: Collaborative House Building with After-Action Explanations}
As we have created an initial version of our HMT testbed and AAE tool, our next step is to validate whether this tool can successfully help humans and machines develop a shared mental model and collaborate better. We provide details regarding our experiment procedure for validating our AAE tool below.

\textbf{High-Level Procedure for User Study:} 
We envision a repeated collaborative house-building task, where users would conduct three iterations of a teaming episode followed by an interaction (or no interaction) with our AAE tool. In each teaming episode, participants would control one Minecraft agent while assisted by a collaborative AI agent. Both team members would have the shared goal of building a structure that was specified to have layers made of different materials. 
Collaboration would be enforced by ensuring that no collaboration would result in elongated build times compared to agents working together. 
To better explicitly measure mental model alignments, we can introduce stochasticity into the AI's perception, decision-making, or action execution capabilities, and see if humans are able to detect why the AI assumes unique behavior or deviates from the expected action. With this experiment with a single independent variable of AAE tool use or not, we would leverage team performance data and fluency metrics \cite{hoffman2019evaluating} to evaluate the quality of HMT. We would also be able to get insight into the specific way the AAE tool was used and receive qualitative feedback on how to better enable the HMT After-Action Review.  

\section{Future Work}

Here, we discuss two promising directions for future work: Agent Code Refinement and Multi-Agent, Multi-Human Collaboration

\begin{enumerate}
    \item Iterative refinement of agent policies: Refining robot policies can be done online (i.e., during a teaming interaction) or offline (i.e., proceeding a teaming interaction). In the offline case, the transcript of the human's interaction with the AAE tool can be used to update the agent's policy. Depending on the AI policy representation, the LLM may be able to directly modify policies in the case of a white-box agent or design a reward function that can be used alongside reinforcement learning in the case of a black-box agent. In the online case, the human may be able to type messages in the testbed chat interface that provide insight into what behaviors the AI is assuming that do not align with their mental model. The transcript of the conversation can be leveraged to update agent policies similar to the offline case.
    \item Multi-Agent, Multi-Human Collaboration: In this paper, we focused on collaboration between one human and one AI agent. However, our HMT testbed and AAE tool can support multiple human and AI agents simultaneously, creating many further multi-agent research opportunities. We hope to explore different avenues of creating collaborative agent teams that can support human teammates.
    
\end{enumerate}






\section{Conclusion}
\label{sec:discussion}




In this work, we have presented two contributions: a Minecraft Human-Machine Teaming testbed and an After-Action Explanation Tool. Our HMT testbed helps to further research in the development and testing of collaborative agents. With our testbed, researchers can now quickly deploy bots to a web-based environment for humans to interact with, addressing a key gap in HMT literature and allowing for human-subjects research without cumbersome setup. 
Our AAE tool furthers research in human-machine mental model alignment.
We include a carefully designed approach to translate HMT gameplay and agent model information into a format ready for ingestion by an LLM and create an interface to enable the replay of a teaming episode and Q$\&$A regarding the teaming interaction via the aforementioned LLM. 

\tiny
\noindent \textbf{Code - }\texttt{\detokenize{https://github.com/MITLL-SMMAAL/pub_HMT_AAEtool_and_testbed}}
\small


\bibliography{bibliography}


\end{document}